\documentclass[preprint,showpacs,preprintnumbers,amsmath,amssymb,superscriptaddress]{revtex4}

\usepackage{graphicx,amsfonts}
\usepackage{dcolumn}
\usepackage{bm}
\usepackage{textcomp}
\begin{document}

\title{Quasi-Dirac neutrinos and solar neutrino data}

 \author{F. Rossi-Torres}%
\email{ftorres@ift.unesp.br}
\affiliation{ Instituto  de F\'\i sica Te\'orica--Universidade Estadual Paulista \\
R. Dr. Bento Teobaldo Ferraz 271, Barra Funda\\ S\~ao Paulo - SP, 01140-070,
Brazil
}
\author{A. C. B. Machado}%
\email{ana@ift.unesp.br}
\affiliation{Centro de Ci\^encias Naturais e Humanas,
Universidade Federal do ABC, Santo Andr\'e-SP, 09210-170\\Brazil
}

\author{V. Pleitez}%
\email{vicente@ift.unesp.br}
\affiliation{ Instituto  de F\'\i sica Te\'orica--Universidade Estadual Paulista \\
R. Dr. Bento Teobaldo Ferraz 271, Barra Funda\\ S\~ao Paulo - SP, 01140-070,
Brazil
}

\date{22/10/2013}
%
\begin{abstract}
We will present an analysis of the solar neutrino data in the context of a $3+1$ quasi-Dirac neutrino
model in which the lepton mixing matrix is given at tree level by the tribimaximal matrix.
When radiative corrections are taken into account, new effects in neutrino oscillations, as $\nu_e \to \nu_s$, could
appear. This oscillation is constrained by the solar neutrino data. In our analysis, we have found an allowed region for
our two free parameters $\epsilon$ and $m_1$. The radiative correction, $\epsilon$, can vary approximately
from $5\times 10^{-9}$ to $10^{-6}$ and the calculated fourth mass eigenstate, $m_4$, varies in the interval 0.01 - 0.2~eV, at 2$\sigma$ level.
 These results are in agreement with the ones presented in the literature in $2+1$ and $2+2$ quasi-Dirac models.
\end{abstract}


\maketitle

\section{Introduction}
\label{intro}

The nature of the neutrino, Dirac or Majorana, is still an open question today. This
notwithstanding, these are not the only options.
For example, neutrinos may be Pseudo-Dirac (PD)~\cite{wolfenstein} or Quasi-Dirac (QD)~\cite{valle}. The former PD
neutrinos arise when two active Majorana neutrinos are mass degenerate. In the QD neutrinos case the mass degeneration occurs with
an active and a sterile neutrino~\cite{nota0}. This sort of neutrinos is called sterile because they do not couple to the $Z$ and neither to the $W$. They only couple, in the context of the known physics, to gravity. In both cases the two Majorana mass degenerate neutrinos
are equivalent to a Dirac one. The QD neutrino also may be generated if the Majorana mass term of the sterile neutrinos is smaller
than the Dirac mass term. In both cases, PD and QD neutrinos,  corrections at the tree level or at the loop level will break the mass
degeneracy. Generally, in models in which one of these options are implemented they are applied to all neutrinos. For instance, all the three
active neutrinos are Dirac, Majorana, PD or QD particles.

An interesting QD situation happens when only one of the active neutrinos together with a sterile one are mass degenerated
at tree level. In this case the three flavor states are, also at tree level, a linear  combination of two massive Majorana
neutrinos and the left component of a Dirac neutrino. This possibility naturally arises when $S_3$ symmetry is
implemented in the neutrino Yukawa interactions~\cite{moha10,schinus,moha11}. In the QD scheme of
Refs.~\cite{moha10,schinus,moha11} the PMNS matrix is, at tree level, the tribimaximal (TBM)~\cite{harrison}
and the scheme is not in agreement with the recent result of a non-zero $\theta_{13}$
angle~\cite{theta13daya,theta13reno,theta13double}. Thus, we can ask ourselves if in the model of Ref.~\cite{schinus}
quantum corrections may induce an appropriate value for that mixing angle. At tree level this is possible if the $S_3$ symmetry
is not implemented in the charged lepton Yukawa interactions and an appropriate value for $\theta_{13}$ is
obtained~\cite{schinus2}. However, quantum corrections imply, in principle, a departure from the TBM that breaks the
mass degeneracy and the PMNS matrix becomes a $4\times4$ matrix. This implies oscillations of active
neutrinos into the sterile one and, for this reason, it is mandatory to analyze how the solar neutrino data constrain
the quantum corrections. The case of a QD with small Majorana masses for the sterile neutrinos was considered in
Ref.~\cite{gouvea}. However, those authors analyzed in detail only QD $2+1$ and $2+2$ schemes. It is not obvious if a
QD $3+1$ scheme~\cite{nota1}, as the present one, satisfies the same constraint as shown in Ref.~\cite{gouvea}.

We know at present more about the parameters of neutrino oscillations and
such knowledge is crucial for us to restructure the Standard Model. For a
recent statistical analysis of all experimental neutrino data available
see~\cite{fogli,valle_stat}. Moreover, since the LEP data, we know that
there are only three active neutrinos~\cite{pdg12}. Thus, an extra neutrino has to be sterile in the sense explained above.
Once sterile neutrinos are added they can be of several types depending on the mass scale related
with them. For a recent review of this sort of neutrinos see Refs.~\cite{whitepaper}.
They may be or not related to some anomalies~\cite{nota2} in neutrino
data~\cite{anomalies_lsnd,anomalies_miniboone,anomalies_reactors,anomalies_detectors}
or with the results of WMAP-7~\cite{wmap7}, which indicates the existence of
four relativistic species ($N_{eff}$).

The main objective of this paper is the following. We will apply the available solar neutrino data in a
more realistic QD $3+1$ case, considering the possibility of electronic neutrinos ($\nu_e$) oscillating to sterile
neutrinos ($\nu_s$). For the statistical analysis, our model
has two parameters: one mass eigenstate ($m_1$) and the radiative
correction ($\epsilon$). This is an important difference if we compare our
analysis with the one made by de Gouv\^ea {\it et al.} in section (III.A 2+1 case)~\cite{gouvea}.
Despite the differences in the model building, we have obtained an allowed
region for $\epsilon$ that is similar to the corresponding values found
in~\cite{gouvea}, which is $\epsilon\approx 10^{-7}$. We stress the fact that the mass splitting of the would-be Dirac neutrinos
do not solve the experimental anomalies presented
in~\cite{anomalies_lsnd,anomalies_miniboone,anomalies_reactors,anomalies_detectors}, however
it is consistent with WMAP-7 results~\cite{wmap7}, since the calculated fourth mass eigenstate, $m_4$, varies in the interval 0.01 - 0.2~eV, at 2$\sigma$ level.

The content of this article is the following: first, in Sec.~\ref{sec:model} we review the basic features of our quasi-Dirac
model, showing its basic structure and possible interactions with the correspondent radiative corrections to the
neutrino mass matrix. In Sec.~\ref{sec:solar_stat} we briefly describe all the solar
neutrino experiments and their respective results. Also we include the method of statistical analysis used. In Sec.~\ref{sec:results} we show in
what condition the model is consistent with solar data, showing the allowed regions in the
parameter space ($m_1,\epsilon$) for our quasi-Dirac model and discussing
the results. Finally, some concluding remarks are presented in Sec.~\ref{sec:final}.

\section{The quasi-Dirac scheme}
\label{sec:model}

Recently it was shown that it is possible that all neutrino flavors are part Dirac and part
Majorana~\cite{moha10,schinus}. The latter occurs because two of the four Majorana neutrinos are mass
degenerate and have opposite parity, so they are equivalent to one Dirac neutrino. As we said before, when these two neutrinos form a Dirac state and they are active, we call
them pseudo-Dirac neutrinos. When there is one active and one sterile, they are called quasi-Dirac neutrinos. The other two have
distinct Majorana masses. In our particular model, we
point out that there are initially three right-handed neutrinos. Two of them are integrated and we
obtain a model QD ``$3+1$'' - three active neutrinos and one sterile.

In this section we are briefly going to describe the construction of our model~(Sec.~\ref{subsec:building}).
In subsection~\ref{subsec:qcorrections}, we show the main interactions that are going to be used to obtain
the radiative corrections for the neutrino masses. These radiative corrections are very important to our
analysis: we study their effects on the break of the degeneracy between the two Majorana neutrinos that
form a Dirac neutrino at tree level. For more details of the model building that we have used here, see~\cite{schinus,schinus2}.

\subsection{The model}
\label{subsec:building}

The model we are going to present here is based on a gauged $B-L$ symmetry with a quasi-Dirac neutrino in
which the right-handed neutrinos carry exotic local $B-L$ charges~\cite{schinus,schinus2}.

When the $S_3$ symmetry is added to the model, the left-handed leptons belong to the reducible triplet
representation ($\textbf{3}=(L_e,L_\mu,L_\tau)$) since all of them have the same $B-L$ charge.

However, unlike the usual case when the three right-handed neutrinos have $L=1$, in this model they have different
$B-L$ charge, so they can transform under $S_3$ only as a singlet $\textbf{1}=n_{\mu R}$ with $B-L=-4$, and a doublet,
$\textbf{2}=(n_{eR},n_{\tau R})$, with $B-L=5$. In the neutrino Yukawa sector, the $S_3$ triplet, $(L_e,L_\mu,L_\tau)$,
can be decomposed into irreducible representations as $\textbf{3}=\textbf{1}+\textbf{2}$, then we can write the singlet
and doublet as follows:
\begin{eqnarray}
 L_2 &=& \frac{1}{\sqrt{3}}(L_e + L_\mu + L_\tau) \sim \textbf{1} \, , \nonumber \\
(L_1,L_3) &=& \left(\frac{1}{\sqrt{6}}(2L_e - L_\mu -
L_\tau),\frac{1}{\sqrt{2}} (L_\mu - L_\tau)\right) \sim \textbf{2}.
\label{s3r}
\end{eqnarray}

The scalar sector has two scalar doublets of $SU(2)$ with weak hypercharge $Y=-1$ that are denoted
by $\Phi_{1,2}=(\varphi^0_{1,2}\,\varphi^-_{1,2})^T$. They are singlets of $S_3$ and we will denote
$\langle\varphi^0_1(\varphi^0_2)\rangle=v_1(v_2)/\sqrt{2}$.
If $n_{\mu R}$ is considered light, but $n_{eR}$ and $n_{\tau R}$ heavy (with masses $m_{n_e}$ and $m_{n_\tau}$, respectively),
we can integrate out the heavy degrees of freedom.
After that, the effective lepton Yukawa interactions are given by a dimension five effective Lagrangian plus
a Dirac mass term, as follows:
\begin{equation}
-\mathcal{L}^{\textrm{eff}}_\nu=h_1\bar{L}_2\Phi_1n_{\mu R}\!+\!
\frac{h^2_2}{m_{n_e}}\,[\overline{(L^c_{1})_R}\,\Phi^*_2][L_{1L}\Phi^*_2]\!+\!
\frac{h^2_3}{m_{n_\tau}}\,[\overline{(L^c_{3})_R}\, \Phi^*_2][L_{3L} \Phi^*_2]
+H.c.,
\label{yuka}
\end{equation}
where the mixing angles in the $(n_{eR},n_{\tau R})$ sector have been absorbed in the dimensionless couplings,
$h_2$ and $h_3$.

From the Yukawa interactions in (\ref{yuka}), we obtain the mass matrix in an appropriate basis~\cite{schinus}, $(\nu_e~\nu_\mu~\nu_\tau~n_\mu^c)_L$. At tree
level the mass matrix is diagonalized by the following $4\times4$ matrix:
\begin{equation}
U_0=\left(\begin{array}{cc}
U_{TBM}&0_{3\times1}\\
0_{1\times3}&1
\end{array}
\right),
\label{utb}
\end{equation}
where $U_{TBM}$ is the tribimaximal matrix and $0$ denote the matrix row or column with entries equal to zero.

Since the model has more interactions than those in the Standard Model, the neutrino mass matrix, when radiative
corrections are taken into account, is not necessarily diagonalized by $U_0$, written in Eq.~(\ref{utb}),
but for $U_0\to U$, where $U$ is now another $4\times4$ matrix which, in principle, is not of the TBM type. In fact the main objective of this paper is verify if the appropriate value for $\theta_{13}$ could arise only from perturbation of the TBM mixing matrix through radiative corrections and keeping agreement with solar neutrino data.
This can be obtained in two different manners: 1) if $(V_{PMNS})$, which is now a $4\times4$ matrix, is such that $(V_{PMNS})_{13}$ has the correct value and $(V_{PMNS})_{14}\sim0$; or 2) both $(V_{PMNS})_{13}$ and $(V_{PMNS})_{14}$  are different from zero but there is a non-negligible
active neutrino into sterile neutrino oscillation. This would imply the disappearance of $\bar\nu_e$ in agreement
with experimental data from Daya Bay, RENO and Double-Chooz.

\subsection{Quantum corrections}
\label{subsec:qcorrections}

\begin{figure}[ht]
\begin{center}
\includegraphics[height=.25\textheight]{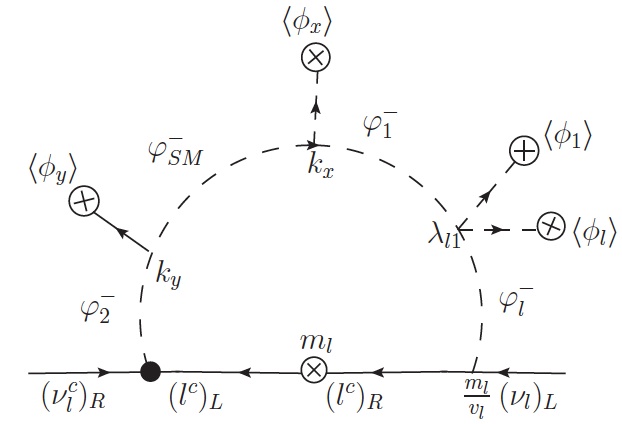}
\caption{\label{fig:loop}
A 1-loop contribution for the Majorana mass matrix in the flavor basis
induced by charged scalars.
The $\bullet$ in the left vertex denotes
interactions in Eq.~(\ref{inter2}).}
\end{center}
\end{figure}

When radiative corrections are taken into account - see Fig.~\ref{fig:loop} - the neutrino mass matrix can be written as
\begin{equation}
M^\nu=M^{0\nu}+\Delta M^\nu,
\label{massa4}
\end{equation}
where $M^{0\nu}$ is the mass matrix at tree level~\cite{schinus2} and $\Delta M^\nu$ arises from 1-loop corrections.
In order to calculate the mass corrections, $\Delta M^\nu$, we have to consider all the Yukawa interactions in the
lepton and scalar sectors. From Eq.~(\ref{yuka}) the scalar-charged lepton interactions are
\begin{eqnarray}
&&-\mathcal{L}^{l\nu \phi}_{CC}=\frac{m_D}{\sqrt{3}v_1}(\bar{e_L}+\bar{\mu}_L+\bar{\tau}_L)
(\nu^c_D)_R\varphi^+_1-\left[\frac{1}{\sqrt6}\frac{m_1}{v^2_2}[2\overline{(e^c)}_R+\overline{(\mu^c)}_R+
\overline{(\tau^c)}_R]\nu_{1L}\right.\nonumber \\&-&
\left.\frac{1}{\sqrt2}\frac{m_3}{v^2_2}[\overline{(\mu^c)}_R-\overline{(\tau^c)}_R]\nu_{3L}\right]
\,\frac{1}{\sqrt2}(v^*_2+\textrm{Re}\varphi^{0*}_2+i\textrm{Im}\varphi^{0*}_2)\varphi^+_2+H.c,
\label{inter2}
\end{eqnarray}
and we have used $\nu^M_{(1,3)L}=\nu_{(1,3)L}$, $\nu^M_{2L}=\nu_{2L}$, $\sqrt{2}(\nu^c_D)_R=n_{\mu R}$. The notation is as
follows: $m_1$ and $m_3$ are the Majorana masses, while $m_D$ is the common mass to the degenerated Majorana neutrinos,
all of them at the tree level. 

The Yukawa interactions from which the charged leptons get mass are mainly the diagonal ones,
\begin{equation}
\mathcal{L}^l_{yukawa}\approx G\sum_l(\bar{\nu}_l \,\bar{l})^\dagger_L\Phi_ll_R+H.c,
\label{inter3}
\end{equation}
where $\Phi_l=(\varphi^+_l\,\varphi^0_l)^T$, and $l=e,\mu,\tau$, where $G$ is a dimensionless constant. The charged lepton mass matrix is
almost diagonal~\cite{schinus}, hence $m_l\approx Gv_l$, and the neutrino interactions with charged leptons are given
by $(m_l/v_l)\bar{\nu}_{lL}l_R\varphi^+_l$. We stress the fact that it is the neutrino flavor
basis that is important here. On the other hand, the scalar potential includes the following
interactions:
\begin{equation}
V(\Phi_1,\Phi_2,\cdots)\propto \lambda_{1l}\Phi^\dagger_l\Phi_1 \Phi^\dagger_1\Phi_l +\lambda_{2l}\Phi^\dagger_l\Phi_2
\Phi^\dagger_2\Phi_l+
k_x\Phi^T_1\varepsilon\Phi_{_{SM}}\phi_x+k_y\Phi^T_2\varepsilon\Phi_{_{SM}}\phi_y+H.c.,
\label{escalares}
\end{equation}
where $l=e,\mu,\tau$. In Eq.~(\ref{escalares}), $\Phi_{_{SM}}$ denotes a scalar doublet with
$Y=+1$ and without $B-L$ charge, and $\phi_x,\phi_y$ are scalars carrying also $B-L$ charges~\cite{schinus}. $k_x$ and $k_y$ are coupling constants with mass dimension and $\varepsilon$ is the antisymmetrical tensor.

With the interactions in Eqs.~(\ref{inter2}), (\ref{inter3}), and (\ref{escalares}), we obtain diagrams like the one
in Fig.~\ref{fig:loop}. As we mentioned before, these sort of diagrams provide corrections to the Majorana masses
for the active neutrinos, i.e., $\overline{(\nu_{aL})^c}\nu_{bL}$. Also corrections to the Dirac mass terms
$\bar{\nu}_{aL}n_{\mu R}$ arise from diagrams similar to the one shown in Fig.~\ref{fig:loop}.
The 1-loop corrections to the neutrino mass matrix ($M^\nu$) in $\chi^\prime_i =N^\prime_{iL}+(N^\prime_{iL})^c$ basis
where $N^\prime_{iL}=(\nu_e\, \nu_\mu\, \nu_\tau\, n^c_\mu )^T_L$ are written in the following:
\begin{equation}
M^{\nu}=m_1 \left(\begin{array}{cccc}
\frac{2}{3}(1-\epsilon_e)&-\frac{1}{3}(1-2 \epsilon_e) \-\-&\- -\frac{1}{3}(1-2 \epsilon_e) \- &\-
\frac{m_D}{\sqrt{3}m_1}(1+\epsilon^\prime_e) \\
-\frac{1}{3}(1-2 \epsilon_\mu)&(\frac{1}{6}+\frac{m_3}{2m_1})(1+2 \epsilon_\mu)\-\-&\-\- (\frac{1}{6}-
\frac{m_3}{2m_1})(1+\epsilon_\mu)\-
&\- \frac{m_D}{\sqrt{3}m_1} (1+\epsilon^\prime_\mu)\\
-\frac{1}{3}(1-\epsilon_\tau)& (\frac{1}{6}-\frac{m_3}{2m_1})(1+\epsilon_\tau)\-\-&\-\-
(\frac{1}{6}+\frac{m_3}{2m_1})(1+\epsilon_\tau)\- &\- \frac{m_D}{\sqrt{3}m_1}(1+\epsilon^\prime_\tau) \\
\frac{m_D}{\sqrt{3}m_1}(1+\epsilon^\prime_e)\-\-&\-\-  \frac{m_D}{\sqrt{3}m_1}(1+\epsilon^\prime_\mu)\-\-
&\-\- \frac{m_D}{\sqrt{3}m_1} (1+\epsilon^\prime_\tau)& 0
\end{array}\right).
\label{massa2}
\end{equation}

From Eq.~(\ref{inter3}), $G=m_e/v_e= m_\mu/v_\mu = m_\tau/v_\tau$, we can express the radiative corrections as:
\begin{eqnarray}
&&\epsilon_l=\frac{1}{8\sqrt2\pi^2}\frac{v_1}{v_2}\lambda_{1l}m^2_l\,A B_l[m^2_l[1-\ln(m^2_l/m^2_{\varphi^+_2})]-
m^2_{\varphi^+_2}+C_l\ln(C_l/m^2_{\varphi^+_2})],\nonumber \\&&
\epsilon^\prime_l=\frac{1}{8\pi^2} \frac{v_2}{v_1}\lambda_{2l}m^2_l\,A B_l[m^2_l[1-\ln(m^2_l/m^2_{\varphi^+_2})]-
m^2_{\varphi^+_2}+C_l\ln(C_l/m^2_{\varphi^+_2})].
\label{epsilons}
\end{eqnarray}
When all $\epsilon$'s and $\epsilon^\prime$'s in Eq.~(\ref{massa2}) are equal to zero, the mass matrix is the same
as the one represented at tree level.

In Eq.~(\ref{epsilons}), $A$, $B_l$ and $C_l$ are given by
\begin{eqnarray}
&&A=\frac{\sqrt{3}\, k_xk_y\langle\phi_x\rangle \langle\phi_y\rangle  }{
 m^6_{\varphi^+_2}},\;\;B_l=\frac{m^6_{\varphi^+_2}}{2(m^2_l-m^2_{\varphi^+_1})(m^2_l-m^2_{\varphi^+_2})(m^2_l-m^2_{\varphi^+_l})
(m^2_l-m^2_{\varphi^0_{SM}})}\nonumber \\ &&
\nonumber \\&&
C_l=m^2_{\varphi^+_l}+m^2_{\varphi^+_2}+m^2_{\varphi^+_1}+m^2_{\varphi^+_3}-5m^2_l.
\label{def}
\end{eqnarray}
In Eq.~(\ref{def}), $m_{\varphi_i^+}$ are the charged scalar masses and $m_l$ are the charged lepton masses.

The general form of the mass matrix in Eq.~(\ref{massa2}) is very complicate to treat, so we will do some approximations in order to simplify our analysis. As we can see in Eqs.~(\ref{def}), we have six dimensionless free  parameters: $\lambda_{1l}$ and $\lambda_{2l}$, where $l=e,\mu,\tau$. Instead of choose the value of each one independently, we use two conditions denoted as CASE A and CASE B, detailed below, and then we have their respective values defined. However, we stress that this numerical choice is not relevant and crucial for our analysis. 

\begin{enumerate}
\item {\bf CASE A}: $\lambda_{1e}m^2_e= \lambda_{1\mu}m^2_\mu=
\lambda_{1\tau}m^2_\tau\equiv M^2_1$ and $\lambda_{2e}m^2_e= \lambda_{2\mu}m^2_\mu=
\lambda_{2\tau}m^2_\tau\equiv M^2_2$. We will also assume that $M_1\approx M_2\sim0.001$ GeV. Note that for this case the value for $\lambda_{1e} \sim \lambda_{2e} < 4$, and the value for the others is even lower.

In this case we have $\epsilon_e=\epsilon_\mu=\epsilon_\tau=\epsilon$ and $\epsilon^\prime_e=\epsilon^\prime_\mu=
\epsilon^\prime_\tau\equiv \epsilon^\prime$ and
\begin{eqnarray}
&&\epsilon\approx \frac{\sqrt3}{16\pi^2}\frac{v_1}{v_2}\frac{k_x\,k_y\langle\phi_x\rangle\langle\phi_y\rangle
M_1^2}{m^2_{\varphi_{SM}}M^4},
\nonumber \\&&
\epsilon^\prime\approx \frac{v^2_2}{\sqrt{2}v^2_1}\,\epsilon.
\label{def2}
\end{eqnarray}
$M$ is a typical mass in the charged scalar sector and $m^2_{\varphi^0_{SM}}$ is the mass square of the
Higgs of the SM. We will use all the scalar masses equal to 125 GeV. In this condition we have
\begin{eqnarray}
\epsilon & = & \frac{5 \times 10^{-21}}{\textrm{GeV}^4}\;  k_x k_y \langle \phi_x\rangle \langle \phi_y\rangle
\nonumber \\
\epsilon^\prime & = & \frac{7 \times 10^{-21}}{\textrm{GeV}^4}\;  k_x k_y \langle\phi_x\rangle \langle \phi_y\rangle
\label{eq12}
\end{eqnarray}
where $k_{x,y}$ and $\phi_{x,y}$ are in GeV units. For $\epsilon \lesssim 1$ we need $k_x k_y \langle\phi_x
\rangle \langle \phi_y\rangle \sim10^{20}\,\textrm{GeV}^4$ which implies four mass scale of the order of 100 TeV, or
at least two masses in the scale of the grand unification. We recall that these dimensional parameters are not related
to the electroweak scale.  Hence, we have put our ignorance about the real values for the parameters in
terms of the scalar sector that is not constrained by the electroweak scale.

\item {\bf CASE B}: $\lambda_{1e}m_e= \lambda_{1\mu}m_\mu=\lambda_{1\tau}m_\tau\equiv M_1$ and $\lambda_{2e}m_e=
\lambda_{2\mu}m_\mu=\lambda_{2\tau}m_\tau\equiv M_2$.
With the same assumption of the CASE A we have
\begin{eqnarray}
\epsilon_l & = & \frac{5 \times 10^{-20}}{\textrm{GeV}^5}\;  k_x k_y \left< \phi_x\right> \left< \phi_y\right> m_l,
\nonumber \\
\epsilon^\prime_l & = & \frac{7 \times 10^{-20}} {\textrm{GeV}^5}\;  k_x k_y \left< \phi_x\right> \left< \phi_y\right> m_l,
\label{etau}
\end{eqnarray}
where $k_{x,y}$ and $\phi_{x,y}$ are in GeV units. In this case we have a certain hierarchy in the radiative corrections
($\epsilon_e\ll \epsilon_\mu\ll \epsilon_\tau$). The radiative corrections $\epsilon_\tau$ and $\epsilon^\prime_\tau$ can be $\lesssim 1$ and
\begin{eqnarray}
&&\epsilon_e =\frac{m_e}{m_\tau} \epsilon_\tau, \quad\epsilon_\mu=\frac{m_\mu}{m_\tau}\,\epsilon_\tau,\nonumber \\&&
\epsilon^\prime_e =\frac{m_e}{m_\tau} \epsilon^\prime_\tau, \quad \epsilon^\prime_\mu=\frac{m_\mu}{m_\tau}\,\epsilon^\prime_\tau,
\label{app2}
\end{eqnarray}
\end{enumerate} 
note that for this case the value for $\lambda_{1e} \sim \lambda_{2e} < 2$, and the value for the others is even lower.

We are going to use both of these approximations in the following analysis. Therefore, we have two main free parameters
which were written in Eq.~(\ref{massa2}): the mass $m_1$ (in eV units) and the radiative corrections (dimensionless) parameter,
$\epsilon$ for the CASE A; and $\epsilon_\tau=\epsilon$ and $\epsilon^\prime_\tau=\epsilon^\prime$ for the CASE B. For this
case, the other $\epsilon$'s and $\epsilon^\prime$'s are calculated by Eq.~(\ref{app2}). We notice that
$\epsilon_\tau\approx\epsilon_\tau^\prime$, then we are going to express our results, for the CASE B, using
$\epsilon_\tau=\epsilon$.

This notwithstanding, it is necessary to analyze how the solar neutrino data constraint the values of $\epsilon$'s since there
are active to sterile neutrino oscillation. This is the issue of the next section.

\section{Solar Neutrinos constraints}
\label{sec:solar_stat}

The detection of neutrinos traveling from the sun has given us a tremendous evidence of neutrino oscillation. We might say
that it was the first time that physicists were doing astronomy with neutrinos and several aspects of the solar behavior
have being observed and understood since then. From Homestake to SNO, nowadays we have a considerable amount
of significant data, which also gives us the opportunity to use this fact to constrain and test the validity of models. This is
exactly what we are going to do: constraining the parameters of the quasi-Dirac model presented in Sec.~\ref{sec:model} and
checking its validity in confrontation with the solar neutrino data. This data is taken from the following experiments:
Homestake~\cite{homestake}, Gallex/GNO~\cite{gallex}, Sage~\cite{sage}, Kamiokande~\cite{kamiokande},
Super-Kamiokande~\cite{sk}, SNO~\cite{sno0} and Borexino~\cite{borexino}. In Sec.~\ref{subsec:review} we are going to present
a small review about these experiments and their main numerical results. In Sec.~\ref{subsec:stat} we present how to treat
the oscillation physics of solar neutrinos and the main points of our statistical analysis. For recent reviews on solar
neutrinos see~\cite{concha,review1,review2}.

\subsection{Experimental data}
\label{subsec:review}

For the statistical analysis, Sec.~\ref{subsec:stat}, we are going to consider the entire set of the solar neutrino data presented
in Table~\ref{tab1}. This table presents each solar neutrino experiment and the measured flux ($\phi_{exp}$). Depending on the
experiment, the flux is measured by charged, neutral current reaction and elastic scattering.

Elastic scattering experiments, $\nu_a + e^- \to \nu_a + e^-$ ($a=e,\mu,\tau$), include Kamiokande~\cite{kamiokande},
Super-Kamiokande~\cite{sk}, Borexino~\cite{borexino,borex_latest}. Kamiokande and Super-Kamiokande detected $^8$B neutrinos
with threshold of 7.5~MeV and~5 MeV, respectively. Borexino, on the other hand, detects neutrinos from the $^7$Be line with
an energy of 0.86~MeV. Recently, Borexino has also measured for the first time the flux of low energy $pep$ neutrinos:
$\phi_{pep}=(1.6\pm0.3)\times 10^8$~cm$^{-2}$s$^{-1}$~\cite{borex_pep}. However, we do not use this value in our analysis.

SNO experiment~\cite{sno0} detects electronic neutrinos in a charged current reaction, $\nu_e + d \to p + p + e^-$
(threshold of 5~MeV). Also, there is the detection of other neutrino flavors by neutral current reaction,
$\nu_a + d \to n + p + \nu_a$ (threshold of 2.225~MeV), and elastic cross section. SNO had three stages and obtained different
fluxes~\cite{sno1,sno2,sno3} shown in Table~\ref{tab1}.

We used also the Homestake experiment~\cite{homestake}, $\nu_e$+$^{37}$Cl $\to$ $^{37}$Ar +e$^-$ (threshold of 0.814~MeV), and
the $^{71}$Ga experiments: GALLEX/GNO~\cite{gallex} and SAGE~\cite{sage} (threshold of 0.233~MeV). In Table~\ref{tab1}, we referred
to all gallium experimental results~\cite{all_gallium}. We notice that they are sensitive to almost the entire neutrino solar spectrum.

\begin{table}[h!]
\centering
\begin{tabular}{|c|c|}
\hline
{\it Experiment} & {\it Experimental Data}\\
\hline
Homestake~\cite{homestake} & $2.56\pm 0.16\pm 0.16$~SNU~~\cite{nota3}\\
\hline
Gallex/GNO and Sage~\cite{all_gallium} & $68.1\pm3.75$~SNU\\
\hline
Kamiokande~\cite{kamiokande} & $\phi_{Kam}=(2.80\pm 0.19\pm 0.33)\times 10^6$~cm$^{-2}$s$^{-1}$\\
\hline
SK~\cite{sk} & $\phi_{SK}=(2.35\pm 0.02\pm 0.08)\times 10^6$~cm$^{-2}$s$^{-1}$\\
\hline
SNO - D$_2$O~\cite{sno1} & $\phi_{CC}=(1.76^{+0.06}_{-0.05}(stat.)^{+0.09}_{-0.09}(syst.))\times 10^6$~cm$^{-2}$s$^{-1}$\\
 &  $\phi_{ES}=(2.39^{+0.24}_{-0.23}(stat.)^{+0.12}_{-0.12}(syst.))\times 10^6$~cm$^{-2}$s$^{-1}$ \\
 &  $\phi_{NC}=(5.09^{+0.44}_{-0.43}(stat.)^{+0.46}_{-0.43}(syst.))\times 10^6$~cm$^{-2}$s$^{-1}$ \\
\hline
SNO - NaCl~\cite{sno2} & $\phi_{CC}=(1.59^{+0.08}_{-0.07}(stat.)^{+0.06}_{-0.08}(syst.))\times 10^6$~cm$^{-2}$s$^{-1}$\\
 &  $\phi_{ES}=(2.21^{+0.31}_{-0.26}(stat.)\pm0.10(syst.))\times 10^6$~cm$^{-2}$s$^{-1}$ \\
 &  $\phi_{NC}=(5.21\pm0.27(stat.)\pm0.38(syst.))\times 10^6$~cm$^{-2}$s$^{-1}$\\
\hline
SNO - $^3$He~\cite{sno3} & $\phi_{CC}=(1.67^{+0.05}_{-0.04}(stat.)^{+0.07}_{-0.08}(syst.))\times 10^6$~cm$^{-2}$s$^{-1}$\\
 &  $\phi_{ES}=(1.77^{+0.24}_{-0.21}(stat.)^{+0.09}_{-0.10}(syst.))\times 10^6$~cm$^{-2}$s$^{-1}$ \\
 &  $\phi_{NC}=(5.54^{+0.33}_{-0.31}(stat.)^{+0.36}_{-0.34}(syst.))\times 10^6$~cm$^{-2}$s$^{-1}$ \\
\hline
Borexino~\cite{borexino} & $\phi=(4.84\pm 0.24)\times 10^9$~cm$^{-2}$s$^{-1}$\\
\hline
\end{tabular}
\caption{Resume of solar neutrino data. Also including the experimental uncertainties. 1 SNU=10$^{-36}$ captures/atom/sec.}
\label{tab1}
\end{table}

\subsection{Analysis}
\label{subsec:stat}

Neutrinos are produced for several thermal nuclear reactions in the center of the sun~\cite{bahcall} and we present them in
Table~\ref{tab:reactions} extracted from~\cite{giuntibook}. The energy of these neutrinos are of a few MeV. To be accurate,
we must treat the center of the sun as a region where the chemical composition modifies itself with the radius. So each
reaction produces a different flux of neutrinos and this changes, as we pointed out, with the position from the center of
the sun. Neutrino sources are: {\it pp, hep, pep, $^{13}N$, $^{15}O$, $^{17}F$, $^{8}B$}, and {\it $^{7}Be$}.
Details of the distribution of the neutrino production as a function of the radius for each of the solar neutrino sources
can be found in~\cite{bahcall} and we used this profile in our work to average the oscillation probabilities, since
detectors only ``see'' these averages.
\begin{table}[h!]
\centering
\begin{tabular}{|c|c|c|c|}
\hline
{\it Source} & {\it Reaction} & {\it Average $\nu$ Energy}~(MeV) & {\it Maximum $\nu$ Energy}~(MeV) \\
\hline
$pp$ & $p+p\to d+e^++\nu_e$ & 0.27 & 0.42\\
\hline
$pep$ & $p+e^- +p\to d+\nu_e$ & 1.44 & 1.44\\
\hline
$hep$ & $^3He+p\to ^4He+e^++\nu_e$ & 9.63 & 18.78\\
\hline
$^7$Be & $e^-+^7Be\to ^7Li+\nu_e$ & 0.86 & 0.86\\
\hline
$^8$B & $^8B\to ^8Be^*+e^++\nu_e$ & 6.74 & 15.00\\
\hline
$^{13}$N & $^{13}N\to ^{13}C+e^++\nu_e$ & 0.71 & 1.19\\
\hline
$^{15}$O & $^{15}O\to ^{15}N+e^++\nu_e$ & 0.99 & 1.73\\
\hline
$^{17}$F & $^{17}F\to ^{17}O+e^++\nu_e$ & 0.99 & 1.74\\
\hline
\end{tabular}
\caption{Sources of solar neutrinos: first column represents the name of the source which produces the electronic neutrino
inside the sun; the second column shows the resume reaction; the third and fourth columns represent, respectively,
the average neutrino energy and the maximum neutrino energy.}
\label{tab:reactions}
\end{table}

After electronic neutrinos ($\nu_e$) are produced by several reactions and in different points of the core of the sun,
they will propagate inside the sun, which has a radius $R_{sun}\approx 6.9\times 10^{10}$~cm. This propagation is described
by the effective Hamiltonian of the system in the flavor state base:
\begin{equation}
H_{eff}(r)=\frac{M^\nu (M^\nu)^\dagger}{2 E} + V(r).
\label{heff}
\end{equation}
We emphasize that $M^\nu=M^\nu(\epsilon,m_1)$ is taken from Eq.~(\ref{massa2}), $V(r)$ is the potential of the
neutrino interaction with the solar environment, $E$ is the $\nu_e$ energy and $r$ is the distance from the center of
the sun. The potential, $V(r)$, can be written as the sum of the charged current and neutral current interaction
($V(r)=V_{cc}(r)+V_{nc}(r)$), which are dependent on the electronic density ($n_e(r)$) and
neutron density ($n_n(r)$) of the environment. Both of these quantities change with the distance from the solar core
and can be written as
\begin{equation}
V(r)=V_{cc}(r)+V_{nc}(r)=\sqrt{2} G_F \left(n_e(r)-\frac{1}{2} n_n(r)\right).
\label{pot}
\end{equation}
The profile of $n_e(r)$ and $n_n(r)$ used in our analysis has been extracted from~\cite{bahcall} and $G_F$
is the Fermi coupling constant.

The survival probability ($P_{ee}$), for each energy and in each point of neutrino production, is calculated from the
amplitude $A_{ee}$, which can be written as:
\begin{eqnarray}
A_{ee}&=&
\left(
\begin{array}{cccc}
1 & 0 & 0 & 0
\end{array}
\right) U_{vac} \times diag(\exp{(-i\Phi^\prime_1)},\exp{(-i\Phi^\prime_2)},\exp{(-i\Phi^\prime_3)},\exp{(-i\Phi^\prime_4)}) \nonumber \\
&\times&
\left(
\begin{array}{cccc}
1 & 0 & 0 & 0\\
0 & \sqrt{1-P_c} & 0 & -\sqrt{P_c}\\
0 & 0 & 1 & 0\\
0 & \sqrt{P_c} & 0 & \sqrt{1-P_c}
\end{array}\right)  \nonumber \\
&\times&
diag(\exp{(-i\Phi_1)},\exp{(-i\Phi_2)},\exp{(-i\Phi_3)},\exp{(-i\Phi_4)})\times
 U_{mat}^\dagger
\left(
\begin{array}{c}
1 \\
0 \\
0 \\
0
\end{array}\right).
\label{aee}
\end{eqnarray}
So the survival probability is written as $P_{ee}=|A_{ee}|^2$. In Eq.~(\ref{aee}), $U_{mat}\equiv U_{mat}(\epsilon,m_1,E)$ is
the matter mixing matrix which diagonalizes the effective Hamiltonian represented by Eq.~(\ref{heff}). The crossing probability,
which will be discussed later, is represented by $P_c$. The matrix $U_{vac}\equiv U_{vac}(\epsilon,m_1)$ is the vacuum mixing
matrix which diagonalizes Eq.~(\ref{heff}) when $V(r)$ is equal to zero (vacuum regime).
If we take $\epsilon=0$, no radiative corrections, for any $m_1$ value, $U_{vac}$ is going to be the tribimaximal mixing
matrix. As the elements of $U_{vac}$, the elements of $U_{mat}$ are modified by the choice of the parameters $\epsilon$ and
$m_1$. When the electronic neutrinos travel to less dense regions of the sun, $U_{mat}\to U_{vac}$. The phases $\Phi_i$
represent the evolution of the mass eigenstates in matter. We express this as $\Phi_i=\int_{r_0}^{R_{sun}}
\mu_i^2(x)/(2E) dx$, where $\mu_i^2(x)$ ($i=1,2,3,4$) is the mass eigenvalue of Eq.~(\ref{heff}) and $r_0$ is the neutrino
point of production. The phase $\Phi_i^\prime$ has a similar meaning as $\Phi_i$, but for the vacuum propagation. We discuss it later in this section.

\begin{figure}[h!tb]
\centering
\includegraphics[height=7.6cm]{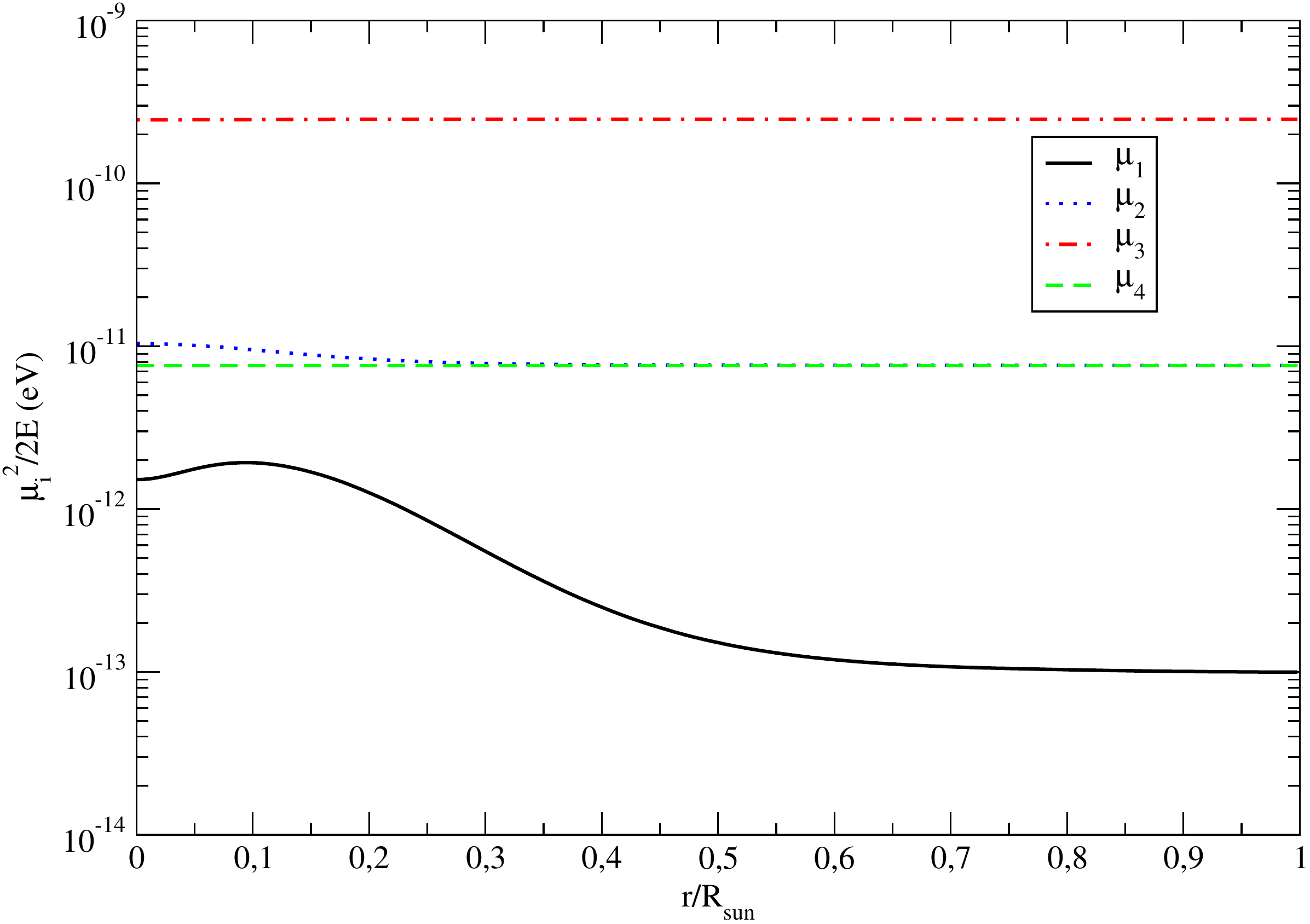}
\caption{Evolution of the mass eigenstates inside the Sun. For this plot we use $\epsilon=1.0\times10^{-3}$, $E=5$~MeV and
$m_1=0.001$~eV. Black solid curve is for $\mu_1$; blue dotted one for $\mu_2$; green dashed curve for $\mu_4$ and red dot-dashed
is for $\mu_3$.}\label{fig:mass}
\end{figure}

In Fig.~\ref{fig:mass}, we show the evolution of the mass eigenstates in the sun for a neutrino with energy $E=5$~MeV,
$m_1=0.001$~eV, $\epsilon=1.0\times 10^{-3}$. The solid black curve represents the mass eigenstate $\mu_1$; dotted blue,
dashed green and dot-dashed red ones represent $\mu_2$, $\mu_4$ and $\mu_3$, respectively~\cite{nota4}. We notice that
$\nu_2$ and $\nu_4$ are practically degenerate, which is the most important characteristic of quasi-Dirac models. It is also
possible to notice that matter can break this degeneracy for very small radius as we can notice in Fig.~\ref{fig:mass}.
However, for very small $\epsilon$, we see that $\nu_2$ and $\nu_4$ are practically degenerate, generating a $\Delta m^2_{42}$
that can be sensible to oscillations: $\Delta m^2_{42} L/(2E)\sim 1$, where $L$ is the Sun-Earth distance, which is about
150~million kilometers.

For instance, in the limit of $\epsilon \to 0$, we recover the original and standard $3\times 3$ situation without the sterile
neutrino presence, where the terms $U_{e1}$ and $U_{e2}$ solve properly the solar neutrino problem: the deficit of $\nu_e$
arriving the Earth. For $\epsilon\ne 0$ and small, it is important to notice that $\nu_2$ and $\nu_4$ will be a coherent mixture
- $(\nu_2 + i\nu_4)/\sqrt{2}$ - and this is an eigenstate of the Hamiltonian in vacuum.

\begin{figure}[h!tb]
\centering
\includegraphics[height=7.6cm]{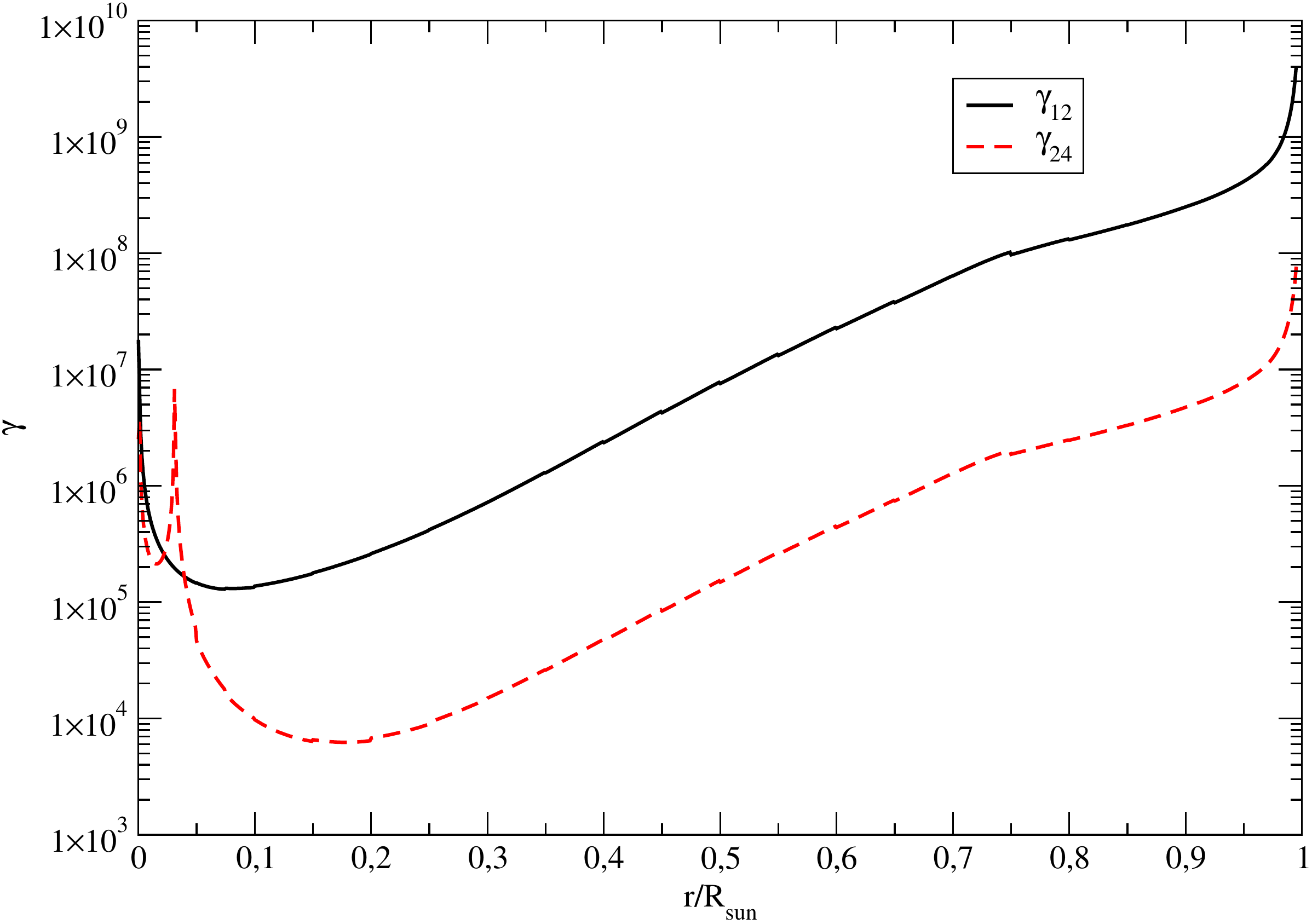}
\caption{Parameter of adiabaticity for the $\nu_1 \to \nu_2$ and $\nu_2 \to \nu_4$ transitions, using $\epsilon=0.8$, $E=5$~MeV, and $m_1=0.001$~eV.}\label{fig:adia}
\end{figure}

We know that the mass eigenstates can feel MSW resonances during the propagation~\cite{msw}. When neutrinos go through the MSW
resonance, the conversion probability is maximal. We remember that conversion probabilities are obtained using the expression
written in Eq.~(\ref{aee}), but changing the position of the number ``1'' of the line vector $(1~0~0~0)$. For example,
$P_{e\mu}$ is obtained using the line vector $(0~1~0~0)$. In principle, we can have resonances among all the mass eigenstates,
however, $\nu_3$ is the heaviest and it will not suffer resonance - its propagation is adiabatic. Also, we can say that the
scale $\Delta m^2_{3i}$, with $i=1,2,4$, can be averaged out. In other words, this mass squared difference scale is not
important for the solar oscillation phenomenon. The moment of the resonance is represented by the matrix that contains $P_c$
in Eq.~(\ref{aee}). This $P_c$ is the crossing probability, which represents the probability of a mass eigenstate $\nu_i$ be
converted to another mass eigenstate $\nu_j$. In the standard neutrino oscillation case, if the propagation is adiabatic, we
must have $P_c=0$. In other words, there is no conversion between two mass eigenstates. In the instantaneous mass basis
($\nu^m_i$, for $i=1,2,3,4$), where $\nu^m$ is the neutrino state in matter, the evolution equation is expressed as:
\begin{equation}
i\frac{d\nu^m}{dx}=\left[\frac{1}{2E} diag(\mu_1^2(x),\mu_2^2(x),\mu_3^2(x),\mu_4^2(x))-iU_m^\dagger(x) \frac{dU_m(x)}{dx}\right],
\label{massinstantaneous}
\end{equation}
where $\mu_i^2(x)$ is the effective mass eigenstate calculated from the eigenvalues of Eq.~(\ref{heff}). If the last term of
Eq.~(\ref{massinstantaneous}) is significant compared with the first one, non-adiabatic transition can happen. The adiabaticity
parameter, represented by the letter $\gamma$, is evaluated at the resonance point, for simplicity, and is defined as
\begin{equation}
\gamma_{ij}=\left|\frac{\frac{(\mu_j^2(x)-\mu_i^2(x))}{2E}}{[U_m^\dagger(x) \frac{dU_m(x)}{dx}]_{ij}}\right|.
\label{adiabaticity}
\end{equation}
When $\gamma_{ij}<<1$ ($\gamma_{ij}>>1$), the propagation is non-adiabatic (adiabatic). Considering $\nu_1$ and $\nu_2$
(or $\nu_4$, since they are practically degenerate), for any values of $\epsilon$ and $m_1$, and evaluating
Eq.~(\ref{adiabaticity}), we conclude that MSW resonance and all the propagation is adiabatic. Actually, for $\epsilon \to 0$,
mixing angles and mass squared differences extracted from our model are very close to the experimental ones~\cite{pdg12}, so we
know from experiments that the propagation is adiabatic. As an example, we can see in Fig.~\ref{fig:adia}, evaluated for
$E=5$~MeV, $\epsilon=0.8$, and $m_1=0.001$~eV, that $\gamma_{12}$ (solid curve) is very large and much greater than 1. Then we can say that
Fig.~\ref{fig:adia} has showed, even for large $\epsilon$, that the value of $\gamma_{12}$ is kept large and then the
propagation remains adiabatic. Even for larger values of $m_1$ we obtain the same pattern and magnitude of $\gamma_{12}$.
We also can say that Fig.~\ref{fig:adia} represents the behavior and magnitude for $\gamma_{14}$. However, since we have a
(quasi-)degenerate state between $\nu_2$ and $\nu_4$, $m_2\approx m_4$, we cannot say that this transition is always
adiabatic. So, $\nu_2 \leftrightarrow \nu_4$ is very dependent on the values of $\epsilon$ and $m_1$. In Fig.~\ref{fig:adia}, $\gamma_{24}$ (dashed curve) is also very high for the $\epsilon$, $m_1$ and $E$ values that we chose. This transition, for this particular choice, is also adiabatic. The modification in the pattern of the curve is related with the modification in the values of the denominator in Eq.~(\ref{adiabaticity}), but this does not modify the adiabatic propagation. When $\epsilon$ becomes smaller, the adiabaticity $\gamma_{24}$ tends to break. We need to compute
Eq.~(\ref{adiabaticity}) and calculate the crossing probability, $P_c$, for this kind of transition.

The crossing probability can be written as~\cite{crossing}
\begin{equation}
P_c=\frac{e^{-\frac{\pi}{2}\gamma |U_{(vac)24}|^2}-e^{-\frac{\pi}{2}\gamma}}{1-e^{-\frac{\pi}{2}\gamma}},
\label{crossing}
\end{equation}
where $U_{(vac)24}$ is the $24$ element of the mixing matrix in vacuum and $\gamma=\gamma_{24}$. If $\gamma$ is large, we have
an adiabatic propagation of $\nu_2$ and $\nu_4$, and they will get out independently of the sun, as distinct mass eigenstates.
Then, in this situation, $P_c=0$. On the other hand, with a very small $\gamma$, which generally happens for very small
$\epsilon$, we have a non-adiabatic propagation of $\nu_2$ and $\nu_4$, and they will get out of the sun, as mentioned before,
as an coherent mixture. Then in this situation, $P_c=0.5$.

After the propagation inside the sun, neutrinos will travel in vacuum, with a phase $\Phi^\prime_i$, $i=1,2,3,4$ - see
Eq.~(\ref{aee}) - where $\Phi^\prime_i=\int m_i^2/(2E)dx$ and $m_i$ is the neutrino mass eigenstate in vacuum. The integration
is taken along all the path from sun to Earth. Only $\Delta m^2_{24}$ is considered, for certain values of $\epsilon$ and $m_1$,
and cannot be averaged out. Other mass squared differences are averaged out since $\Delta m^2 L/(2E)>>1$. It is important to
stress out that we are ignoring Earth matter effects. An extended analysis involving four neutrino families with a great variety
of $\Delta m^2_{14}$ and mixing angles was done in~\cite{cirelli}.

The general expression of the expected event rate in the presence of oscillations in experiment $j$ in the four neutrino
framework is given by $R_j^{th}$:
\begin{equation}
R_j^{th}=\sum_{k=1,8}\phi_k\int dE \lambda_k(E)\times [\sigma_{e,i}(E)\langle P_{ee} \rangle+\sigma_{x,i}(E)(1-\langle P_{ee}
\rangle-\langle P_{ea} \rangle)],
\label{rate}
\end{equation}
where $E$ is the neutrino energy, $\phi_k$ and $\lambda_k$ are, respectively, the total neutrino flux and the neutrino energy
spectrum normalized to one from the solar nuclear reaction $k$ with normalization given by the model BS05(OP) in~\cite{bahcall}
- see Table~\ref{tab:reactions}. In Eq.~(\ref{rate}), $\sigma_{e,i}$ ($\sigma_{x,i}$) is the $\nu_e$ ($\nu_x$,$x=\mu,\tau$)
interaction cross section in the Standard Model with the target corresponding to experiment $j$, $\langle P_{ee} \rangle$ is
the average survival probability in the production point, $\langle P_{ea} \rangle$ and $\langle P_{es} \rangle$ are, respectively,
the average conversion probability in the production point of $\nu_e \to \nu_a$ ($a=\mu,\tau$) and $\nu_e \to \nu_s$.

The $\chi^2$ test is calculated by
\begin{equation}
\chi^2=\sum_j\frac{(R_{th}^j-R_{exp}^j)^2}{\sigma^2_j},
\label{chi}
\end{equation}
where $R_{exp}^j$ is the experimental rate for $j$-experiment - see Table~\ref{tab1}. Generally, the rate is defined as
$R=\phi/\phi_{SSM}$, where $\phi_{SSM}$ is the total flux of the solar standard model extracted from the model
BS05(OP)~\cite{bahcall}: $R_{th}$ corresponds to a $\phi_{th}$ that represents the oscillated flux, which is related to the
parameters of our model and evaluated using Eq.~(\ref{rate}); $R_{exp}$ is based on the flux $\phi_{exp}$, which is the
experimental value extracted from Table~\ref{tab1}. The $\chi^2$ is calculated for each set of $\epsilon$ and $m_1$. Note
in Eq.~(\ref{chi}) that $\sigma^2_j$ is the error, which takes into account the experimental error of a particular experiment
and errors associated with the flux expectations in BS05(OP).

\section{Results}
\label{sec:results}

\begin{figure}[h!tb]
\centering
\includegraphics[height=7.6cm]{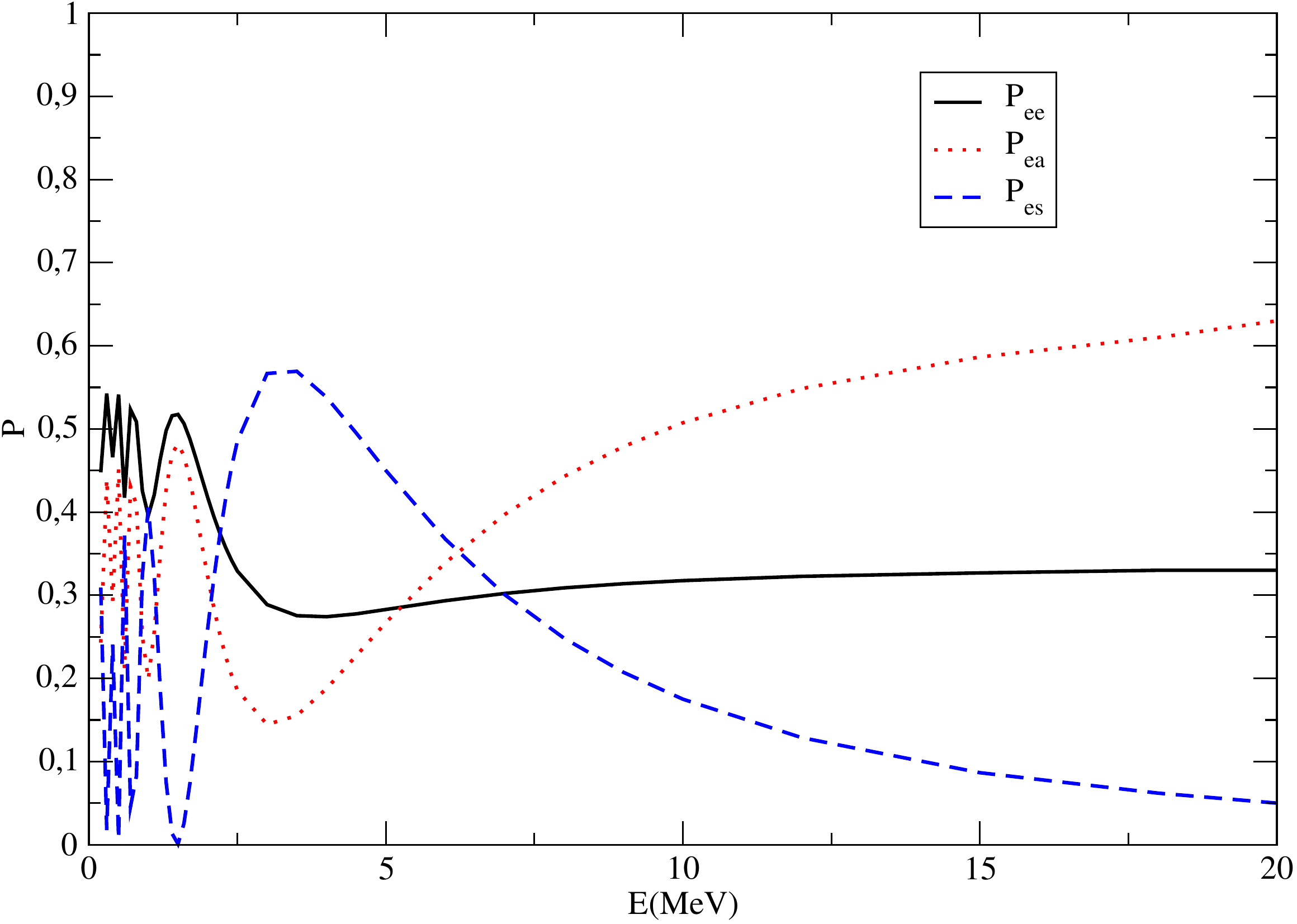}
\caption{Survival probabilities ($P_{ee}$ - solid black curve) and conversion probabilities ($P_{ea}$ - dotted red curve, -
and $P_{es}$ - dashed blue curve) for the CASE A. All the curves correspond to $\epsilon=5.0\times 10^{-7}$ and $m_1=0.003$~eV.
We do not show these probability curves for the CASE B, since their behavior is very similar and there are very small
differences.}\label{pee1}
\end{figure}

From Eq.~(\ref{aee}), we plot the probabilities for two sets of parameters $(m_1,\epsilon)$ for the CASE A. In Fig.~\ref{pee1},
we used as input $\epsilon=5.0\times 10^{-7}$ and $m_1=0.003$~eV. The masses $m_2$ and $m_3$, which appear in Eq.~(\ref{massa2}),
are calculated. To maintain hierarchy, for which we choose the normal one for simplicity, the masses $m_2$ and $m_3$ will be written
as follows: $m_2=\sqrt{\Delta m^2_{sun}+m_1^2}$ and $m_3=\sqrt{\Delta m^2_{atm}+m_1^2}$. The values of $\Delta m^2_{sun}=7.58
\times 10^{-5}$~eV$^2$ and $\Delta m^2_{atm}=2.35\times 10^{-3}$~eV$^2$ are the best-fit values at 1$\sigma$ taken from~\cite{pdg12}.

After introduce these values of masses in Eq.~(\ref{massa2}), we evaluate the neutrino evolution Hamiltonian in matter,
Eq.~(\ref{heff}), calculating the new mass eigenvalues and diagonalizing it to obtain the new mixing matrix. It is important to
notice that for each set ($\epsilon$,$m_1$) we are going to have new elements of the mixing matrix and new mass eigenstates either
in vacuum and in matter. Then we evaluate the probabilities using Eq.~(\ref{aee}) and average them in the region of production. In
Fig.~\ref{pee1}, the survival probability, $P_{ee}$, is represented by the solid black curve. Conversion probabilities are
represented in the following way: $P_{ea}$ ($a=\mu,\tau$) is the dotted red curve and $P_{es}$ is the dashed blue curve.

\begin{figure}[h!tb]
\centering
\includegraphics[height=7.6cm]{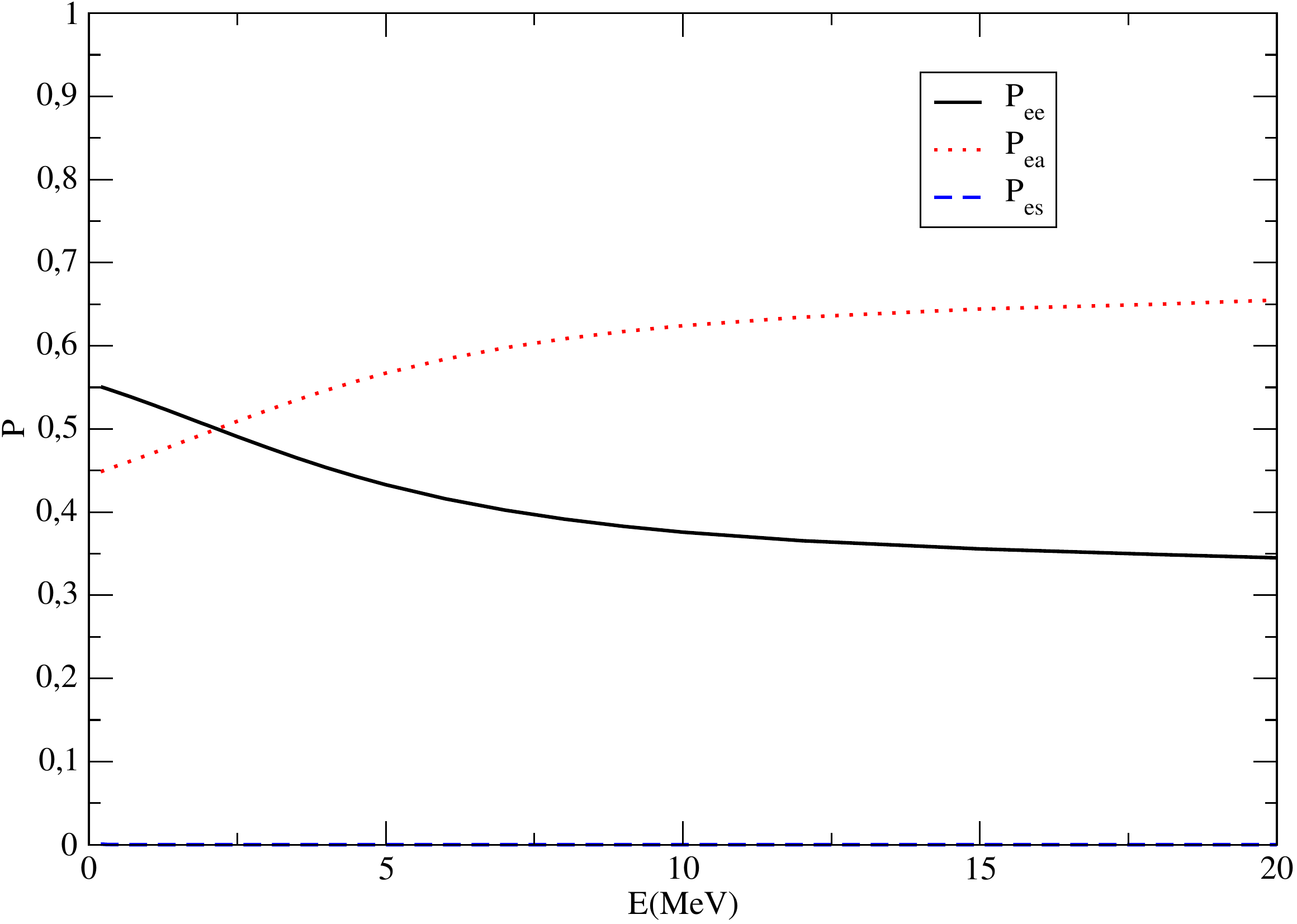}
\caption{Survival probabilities ($P_{ee}$ - solid black curve) and conversion probabilities ($P_{ea}$ - dotted red curve, -
and $P_{es}$ - dashed blue curve) for the CASE A. All the curves correspond to $\epsilon=1.0\times 10^{-9}$ and $m_1=0.003$~eV.
Similar results are obtained in CASE B.}\label{pee2}
\end{figure}

In Fig.~\ref{pee2} we have plotted the probabilities for the CASE A for $\epsilon=1.0\times 10^{-9}$ and $m_1=0.003$~eV. Notation
and representation of the curves are the same as in Fig.~\ref{pee1}.

We notice, for example, for small $\epsilon$, such as $\epsilon=1.0\times 10^{-9}$ in Fig.~\ref{pee2}, that the quasi-Dirac
situation mimics the standard one, since we do not see a significant conversion to the sterile neutrino flavor. So, when
$\epsilon \to 0$, or simply to very small values, we approach the traditional solar neutrino solution. However, for higher
$\epsilon$, such as we saw in Fig.~\ref{pee1} ($\epsilon=1\times 10^{-7}$), conversion to sterile neutrinos can be significant
for the entire neutrino spectrum. If we get an even higher $\epsilon$, we will see an even larger oscillation pattern of $P_{es}$.
This also happens for the other channels of oscillations ($\nu_e \to \nu_{\mu,\tau}$). That is because with a larger radiative
correction, $\epsilon$, we get $\Delta m^2_{ij} L/(2E)>>1$. The phenomenological effect of $m_1$ is very similar. For large
values of $m_1$, if we fix $\epsilon \ne 0$, we are going to have more oscillation if we compare to the situation with a smaller
$m_1$.

One of the main sources of neutrinos, considering SNO and SK as experiments, is the $^8$B. For energies above a few MeV, SNO
and SK reveal that $P_{ee}\approx 0.3$ and also $P_{ea}\approx 0.7$. This is a very strong constraint. For energies below 1~MeV
or so, the constraints come mainly from Borexino, Homestake, and the gallium experiments. Borexino imposes $P_{ee}\approx 0.51$.
Then if we have larger values of $P_{es}$, $P_{ee}$ must be higher to compensate the disappearance of active neutrinos ($\nu_\mu$
or $\nu_\tau$) that would arrive in Earth detectors. For even lower neutrino energies, mainly of the $pp$ chain, gallium
experiments impose $P_{ee}\approx 0.5$. So any modification on $P_{es}$ for the set of parameters $\epsilon, m_1$ has to be
compensated by $P_{ee}$, especially in the high energy part of the spectrum.

We do not show the plots for the CASE B, because the behavior and the pattern of the curves are very similar.

Next, we proceed with a $\chi^2$~fit to the data. This will be used to constrain our model for both CASES A and B. We can define
$\Delta \chi^2=\chi^2(\epsilon,m_1)-\chi^2(\epsilon=0)$, where $\chi^2(\epsilon=0)$ is valid for any value of $m_1$, since
$\epsilon=0$ represents the standard situation and solar neutrino experiments are sensitive only to the mass squared difference
and not to the absolute value of neutrino masses. For $\epsilon\ne 0$, we choose $m_1$ to vary from 0.001~eV to 1~eV. We remember
that Katrin will impose a superior limit on neutrino mass of about 0.2~eV~\cite{katrin}.

\begin{figure}[h!tb]
\centering
\includegraphics[height=7.5cm]{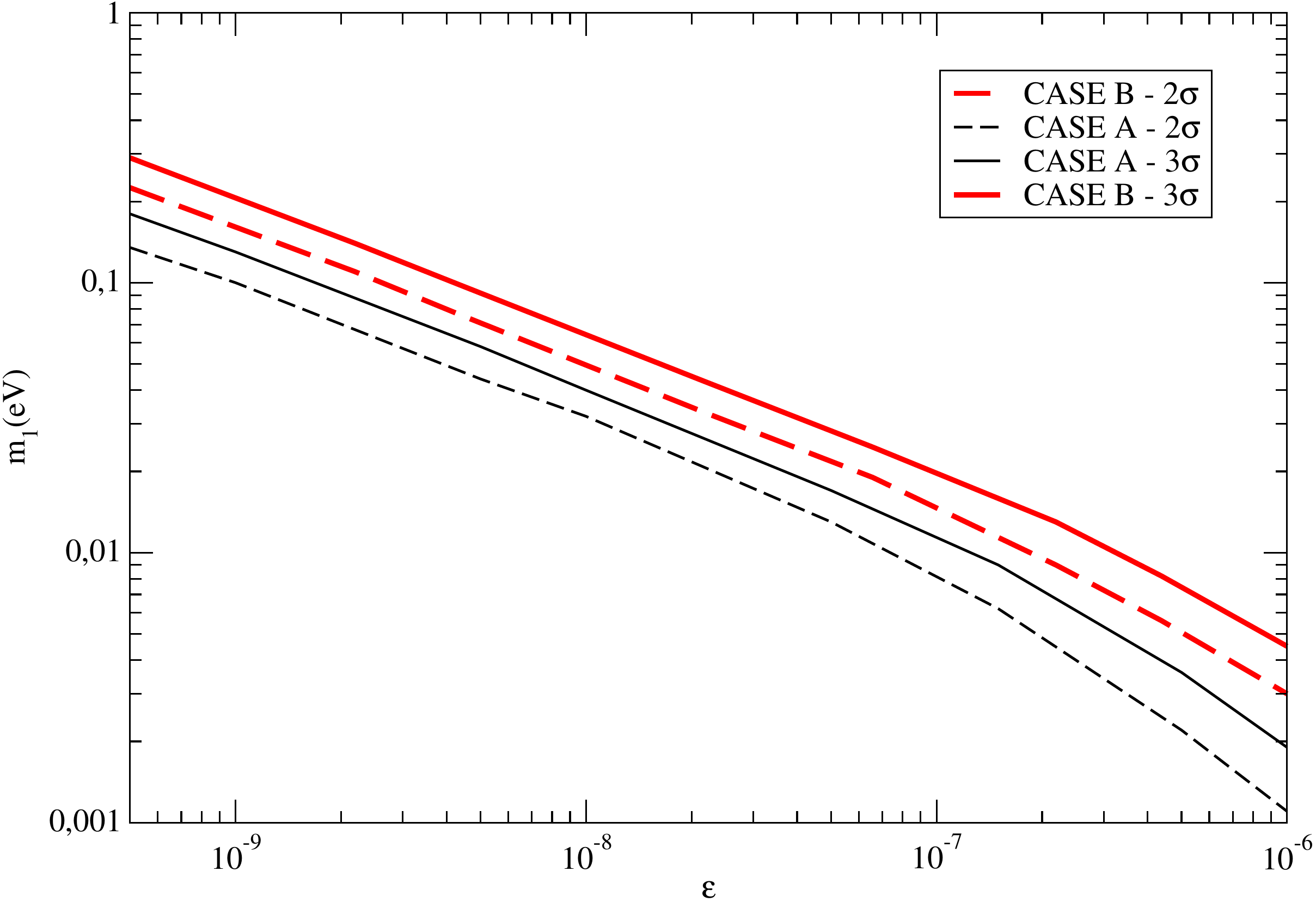}
\caption{Allowed regions for the parameters $\epsilon$ and $m_1$. Thinner curves (black curves) represent the CASE A. Thicker
curves (red curves) represent the CASE B.}\label{todos_m1}
\end{figure}

\begin{figure}[h!tb]
\centering
\includegraphics[height=7.5cm]{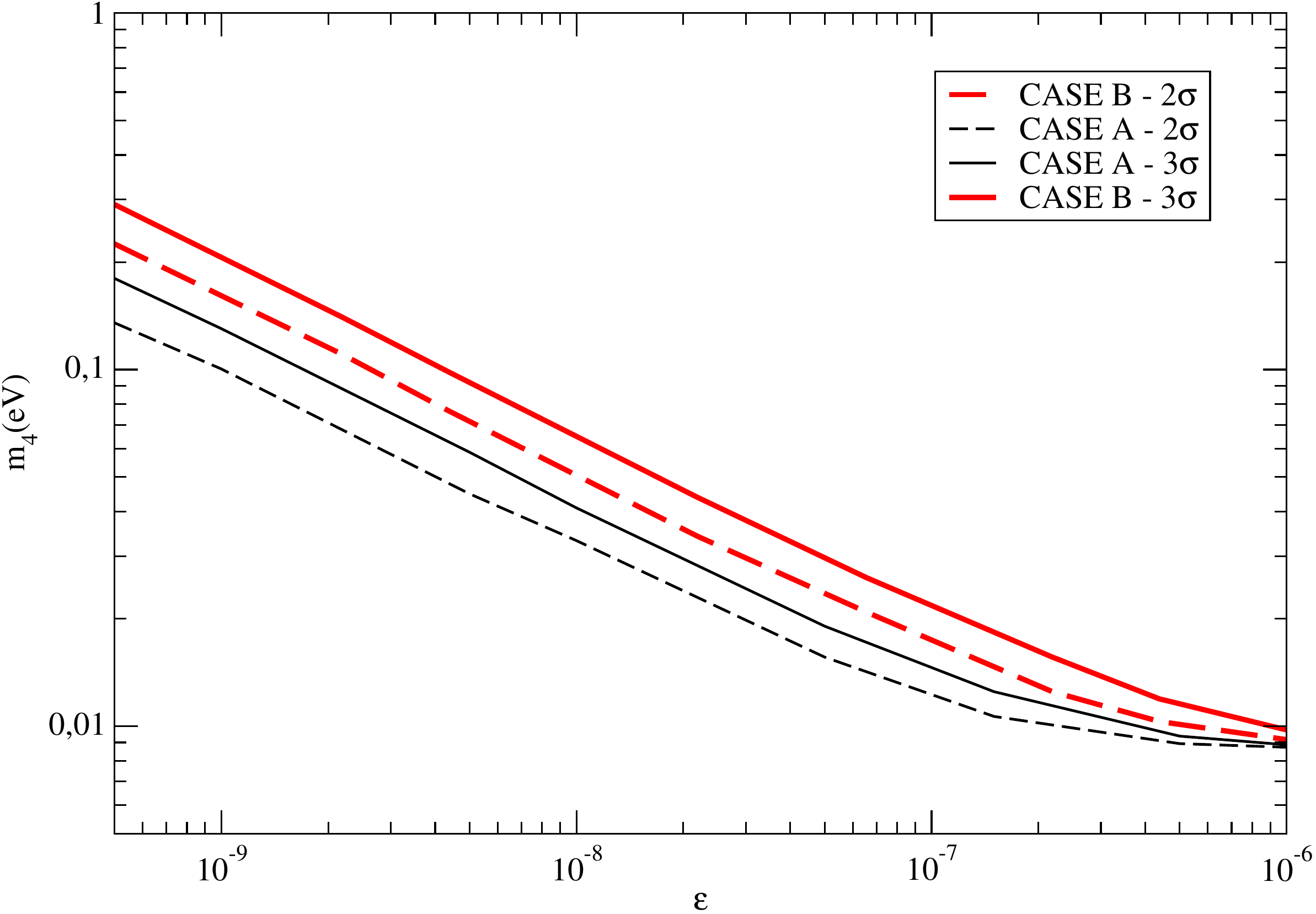}
\caption{Allowed regions for the parameters $\epsilon$ and $m_4$. Thinner curves (black curves) represent the CASE A. Thicker
curves (red curves) represent the CASE B.}\label{todos_m4}
\end{figure}

In Fig.~\ref{todos_m1}, we present the allowed region for the CASE A and CASE B together. Below the curves are the allowed regions.
The dashed curves are the 2$\sigma$ allowed region for the parameters $\epsilon$ and $m_1$. The solid curves are the 3$\sigma$
allowed region. Thinner curves (black ones) represent the CASE A and thicker curves (red ones) represent the CASE B.
In Fig.~\ref{todos_m4} we have the same analysis, now for $m_4$ values. We notice, as shown in both Fig.~\ref{todos_m1} and Fig.~\ref{todos_m4},
that the value of the scale of $m_1$ is very similar to $m_4$, since $m_4\sim m_2$, and $\Delta m^2_{21}\sim \Delta m^2_{41}
\approx 7.5\times 10^{-5}$~eV$^2$ for small $\epsilon$. As the radiative correction ($\epsilon$) grows, $m_1$ ($m_4$) has to
diminish to maintain the $\chi^2$. They have, in some sense, a compensatory behavior when considered together. We remember for 
CASE B that $\epsilon=\epsilon_\tau$ ($\epsilon^\prime=\epsilon^\prime_\tau$), and $\epsilon$ and $\epsilon^\prime$ are related by
Eq.~(\ref{etau}).

In Fig.~\ref{todos_m1} and Fig.~\ref{todos_m4}, we have plotted all the curves together to make evident the difference between
CASE A and CASE B. We notice that there is a very small difference between these two approximations, which would be evidence that
it is almost impossible to distinguish between them.

Our results are similar to the ones found by de Gouv\^ea {\it et al.}~\cite{gouvea}. They found in the $2+1$ case (two active
neutrinos + one sterile) that $\epsilon<2.0\times 10^{-7}$ for 3$\sigma$ and $\epsilon<1.2\times 10^{-7}$ for 2$\sigma$. Their
model has only the $\epsilon$ parameter, however, our model possesses two parameters. Despite this fact, as we can see in
Fig.~\ref{todos_m4}, we also have found similar values of $\epsilon$ when 0.01~eV$<m_4<$0.2~eV at 2$\sigma$ level. The important
characteristic of the model, quasi-degeneracy with the $m_2$ state ($\Delta m^2_{24}$ much smaller than the other ones), is still
maintained even for these $m_4$ values of masses. In both cases, the mixing matrix $U$ after radiative corrections and solar neutrino data analysis remains the TBM one for all practical purposes, represented by Eq.~(\ref{utb}).

\section{Conclusions}
\label{sec:final}

In this paper we have analyzed the model with a quasi-Dirac neutrino put
forward in Refs. \cite{schinus,schinus2} using the solar neutrino data.
This is possible because when radiative corrections are included in the
neutrino mass matrix the oscillation channel $\nu_e\to \nu_s$ is open.
However,
we have got the result that, even in this case, the quasi-Dirac neutrino remains, for
all practical proposes, a Dirac one, i.e., $m_2\approx m_4$.
Our model has two parameters, the radiative correction $\epsilon$ and the
input mass $m_1$, which is the small one considering, for simplicity, the
normal hierarchy.
We have found allowed regions, shown in Fig.~\ref{todos_m4}, in which $\epsilon$ can
vary approximately from $5\times 10^{-9}$ to $10^{-6}$ and the $m_4$ mass
varying from 0.01~eV (0.01~eV) to 0.2~eV (0.3~eV) at
2$\sigma$ (3$\sigma$) level. Using Eq.~(\ref{eq12}), this implies $10^{12}$~GeV$^4$ $\lesssim k_x k_y \langle \phi_x \rangle \langle\phi_y \rangle \lesssim$ 10$^{14}$~GeV$^4$, which means that we have four mass scales between $\sim$1~TeV and $\sim$3~TeV. The $m_4$ values are compatible with the most conservative limit of the sum of neutrino masses
($\sum_{m_\nu}<1.3$~eV (95\%)) of WMAP-7~\cite{wmap7}. In Ref.~\cite{gouvea}, which describes a $2+1$ model- two active
neutrinos plus one sterile - it was obtained $\epsilon<(1.2,2.0)\times 10^{-7}$ at two and three sigma
level, respectively. Note that in our case the four masses belong to the interval with order of magnitude $\sim (10^{-3}$
 -- $10^{-1})$~eV.

Summarizing, even with radiative corrections are considered, the mixing
matrix in the lepton sector continues to be the tribimaximal one.
It means that in these conditions the model cannot explain the
disappearance of $\bar\nu_e$ observed by several experiments and, when
interpreted in a three
active neutrino scenario, it implies a non-zero
$\theta_{13}$~\cite{theta13daya,theta13reno,theta13double}. Hence, the
only way to obtain a realistic $PMNS$ mixing matrix is
by considering a non-diagonal charged lepton mass matrix as has been put
forward in Ref.~\cite{schinus2}. This
will introduce contributions with crossed masses: $m_em_\mu$, etc., but none of them are as important
like the term proportional to $m^2_\tau$. By doing this our results will not be significantly modified.

\acknowledgements
FR-T would like to thank CNPq for the financial support, ACBM would like to thank CAPES for the financial support and
VP would like to thank CNPq  and FAPESP for partial financial support. One of us (FR-T) would like to thank the important discussions
and suggestions made by, A. M. Gago, O. L. G. Peres, M. M. Guzzo and P. C. de Holanda during the elaboration of this manuscript.



\begin{thebibliography}{99}
\bibitem{nota0} Usually both terms are used as synonymous in literature. However, we prefer the present
notation which allows one to distinguish different mechanism in model building.
\bibitem{nota1}Traditionally in the literature the 3+1 scheme refers to three active neutrinos and one sterile neutrino which is
necessarily heavier than the others. We stress the fact that our ``$3+1$'' scheme is a quasi-Dirac one.
\bibitem{nota2}When we refer to neutrino anomalies we are talking about the need of more families to explain the results of,
for example, LSND and MiniBooNE experiments and others that require $\Delta m^2>0.1$~eV$^2$.
\bibitem{nota4}The use of the Greek letter $\mu_i$ is generally used to represent the mass eigenstates in matter;
$m_i$, on the other hand, is usually used to represent the neutrino mass eigenstates in vacuum.
\end{thebibliography}
\end{document}